\documentclass[journal,twoside,web]{ieeecolor}
\usepackage{generic}
\usepackage[T1]{fontenc}
\usepackage[utf8]{inputenc}
\usepackage{cite}
\usepackage{amsmath,amssymb,amsfonts}
\usepackage{graphicx}
\usepackage{xcolor}
\usepackage{url}
\usepackage{booktabs,array}
\usepackage{textcomp}
\usepackage{placeins}
\usepackage{newunicodechar}
\newunicodechar{ρ}{\ensuremath{\rho}}
\newunicodechar{α}{\ensuremath{\alpha}}
\newunicodechar{ω}{\ensuremath{\omega}}
\newunicodechar{θ}{\ensuremath{\theta}}
\newunicodechar{·}{\ensuremath{\cdot}}
\newunicodechar{–}{--}
\newunicodechar{—}{---}

\usepackage{hyperref}
\hypersetup{
  colorlinks=true,
  linkcolor=nblue,
  citecolor=blue,
  urlcolor=blue,
  pdfborder={0 0 0}
}

\def\BibTeX{{\rm B\kern-.05em{\sc i\kern-.025em b}\kern-.08em
T\kern-.1667em\lower.7ex\hbox{E}\kern-.125emX}}
\def\journalname{Journal}
\begin{document}
\title{Filamentary Transport and Thermoelectric Effects in Mushroom Phase Change Memory Cells}
\author{Md Samzid Bin Hafiz, Helena Silva, and Ali Gokirmak
\thanks{Authors are with the University of Connecticut, CT 06269 USA (e-mail: ali.gokirmak@uconn.edu). }}

\bstctlcite{IEEEexample:BSTcontrol}
\maketitle
\begin{abstract}
We performed a 2D finite-element electro-thermal computational study of thermoelectric effects and filamentary electronic transport in Ge$_2$Sb$_2$Te$_5$ mushroom phase change memory cells during Reset and Set operations, accounting for spatial activation energy variations in amorphous Ge$_2$Sb$_2$Te$_5$ and phase-change dynamics. Reset operations with current going from the top electrode to the narrow (4 nm) bottom electrode requires $\sim$ 3x less energy and power, and $\sim$ 2x lower current to achieve the same Reset resistance, compared to the opposite polarity, due to thermoelectric effects. Filamentary conduction, electrical breakdown, thermal runaway and local crystallization of amorphous Ge$_2$Sb$_2$Te$_5$ depend on current polarity and thermal boundary conditions, and determine the location, shape and volume of the programming region, which may be significantly smaller than the semi-cylindrical mushroom region. The programming volume does not scale with contact dimensions larger than 10 nm. Larger contact areas introduce increased device-to-device and cycle-to-cycle variability due to filamentary conduction but are expected to lead to higher reliability and endurance.
\end{abstract}


\section{Introduction}
\normalcolor

\IEEEPARstart{I}{N} the era of artificial intelligence, big data and neuromorphic computing, the demand for high-speed, energy efficient and scalable memory technologies has surged beyond the capabilities of conventional memory \cite{zhang2019designing}. The performance bottlenecks in von Neumann architectures largely stem from the latency gap between the CPU, volatile DRAM, and nonvolatile flash memory. Non-volatile resistive memory (RRAM) technologies such as memristors, magnetic RAM (MRAM) and phase change memory (PCM) emerged as possible candidates to bridge this gap. PCM is the most mature among these \cite{burr2016recent} and it is in high volume production \cite{redaelli2022material,cappelletti2020phase}.

The programming region of PCM devices is composed of a phase change material, which can be recursively switched between its crystalline (low-resistance) and amorphous (high-resistance) phases by self-heating with short duration voltage pulses \cite{faraclas2011modeling}. The programming region is amorphized by heating above melting temperature ($T_{\mathrm{melt}}$), followed by rapid quenching, and recrystallized by heating above glass-transition temperature ($T_{\mathrm{glass}}$).

\begin{figure}[!tbp]
\centering
\includegraphics[width=\columnwidth]{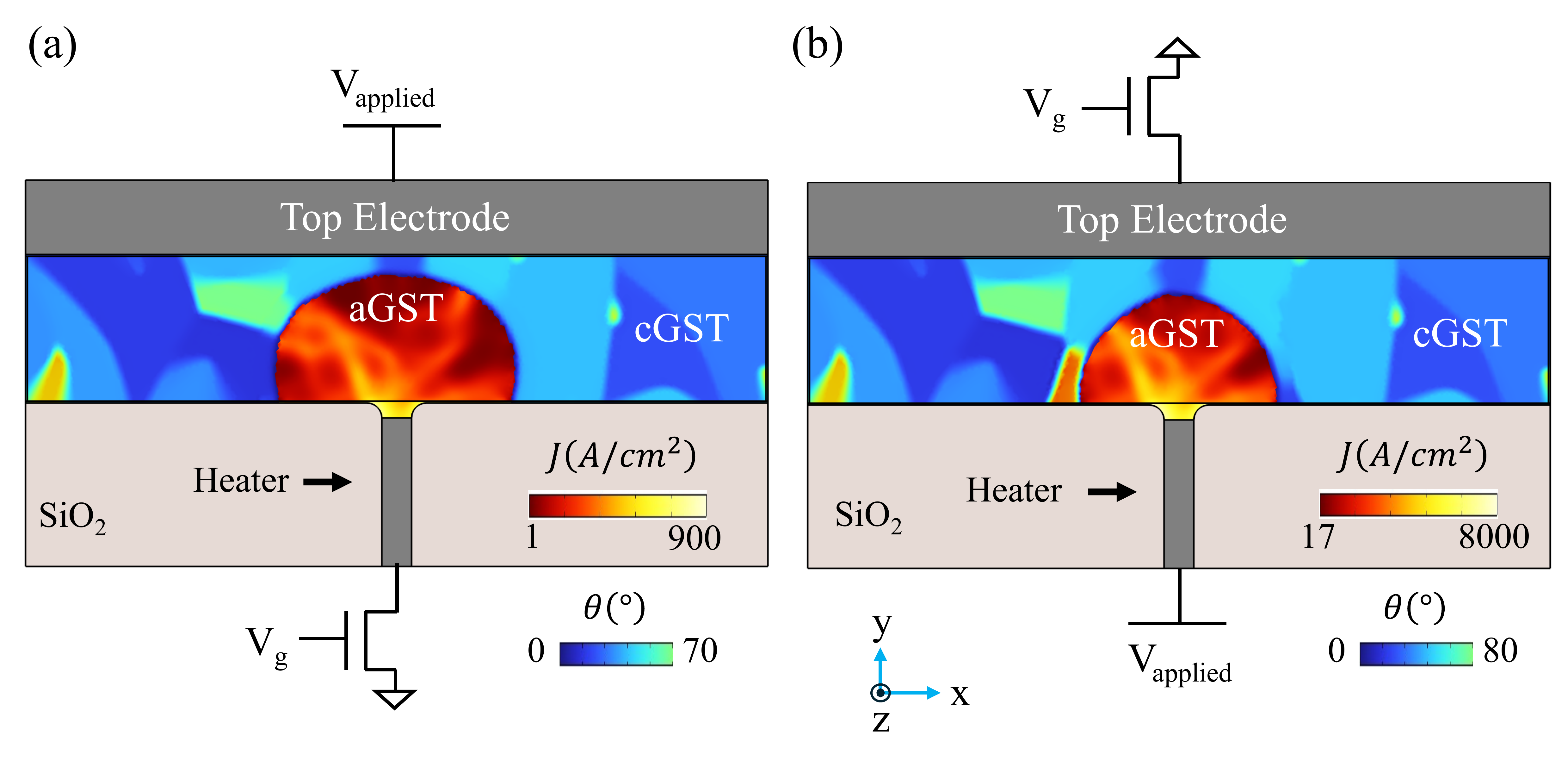}
\caption{PCM cell with a MOSFET in series for (a) top and (b) bottom bias polarity. Crystal orientation angles for crystalline phase and current density profiles for amorphous phase of PCM are shown. $W_{\mathrm{heater}}=4~\mathrm{nm}$.}
\label{fig:circuit.pdf}
\end{figure}

PCM can be SET and Reset with $\sim 1$--$100~\mathrm{ns}$ voltage pulses in the $\sim 1~\mathrm{V}$ range, achieving large resistivity contrast ($10^2$--$10^4\times$) in a CMOS compatible resistance range ($\mathrm{k}\Omega$--$\mathrm{M}\Omega$) and $>10$ years retention at CPU operating temperatures \cite{hafiz2025modeling,raoux2014phase}. PCM can be Reset, Set and read using unipolar voltage pulses (without switching the current direction) and is scalable down to $\sim 5~\mathrm{nm}$, unlike memristors and MRAM \cite{raoux2008phase}. Read operation is non-destructive, performed using small voltage pulses ($\sim 0.1~\mathrm{V}$), and can be very fast ($<1~\mathrm{ns}$). Current, power and energy requirements for Reset and Set operations scale with phase change material and device size. Ge$_2$Sb$_2$Te$_5$ (GST) is widely adopted as the phase change material due to its low thermal conductivity, and suitable melting and crystallization kinetics \cite{faraclas2011modeling,yu2023observation}. Germanium-rich GST, with a higher $T_{\mathrm{glass}}$, is used for high-temperature applications \cite{redaelli2022material}. Cross-point (sandwich-cell) PCM devices integrated with Ovonic Threshold Switch (OTS) access devices have been mass-produced using $20~\mathrm{nm}$ technology as 3D Xpoint \cite{hafiz2025modeling,burr2016recent,cheng20193d}. Mushroom cells, constructed with a common top contact and small bottom contacts, have programming regions surrounded by crystalline phase change material and do not require nucleation for crystallization, leading to higher-speed, and offer the flexibility to use side-wall processes to fabricate very narrow bottom contacts ($W_{\mathrm{heater}}\sim 2$--$3~\mathrm{nm}$) \cite{ielmini2018phase}. If a side-wall process is used for contact definition, rows of devices are formed by lithographically defining and etching the common top-contact and phase change layer, defining the width of the phase change material (out of plane dimension of Fig.~\ref{fig:circuit.pdf}), and the programming regions form as semi-cylinders \cite{raoux2008phase}.

PCM cells experience high temperatures ($T\sim900~\mathrm{K}$), extreme current densities ($J\sim100~\mathrm{MA/cm^2}$) and thermal gradients ($\sim 50~\mathrm{K/nm}$) which make the contribution of thermoelectric effects very significant. The Set operation relies on electrical breakdown of the amorphous region, giving rise to highly non-linear current-voltage characteristics that are strong functions of temperature \cite{ielmini2008threshold}. Hence, understanding the electronic conduction within the amorphous region and at its interfaces is critical for device and waveform design.

Trap-assisted tunneling, Poole--Frenkel transport, and field induced delocalized tail states models have been proposed as electronic conduction mechanisms within amorphous phase change materials \cite{pan2014recent,nardone2012electrical}. The significant band offsets at the amorphous phase change materials' interfaces are expected to lead to energy barriers for electrons and holes. Local variations in the disordered amorphous phase give rise to significant variations in current density, local self-heating, filamentary transport and thermal runaway \cite{hafiz2025modeling}. In our recent modeling study, we have incorporated local variations in amorphous (a-) GST properties, going beyond the commonly used effective-media approximation for a-GST, and demonstrated filament formation in a sandwich-cell structure \cite{hafiz2025modeling}. In this work, we investigate filamentary conduction, impact of thermoelectric effects and the phase change dynamics during Reset and Set operations of GST mushroom cells \cite{ielmini2008threshold}.

\section{Modeling Framework}
\normalcolor

\begin{figure}[!t]
\centering
\includegraphics[width=\columnwidth]{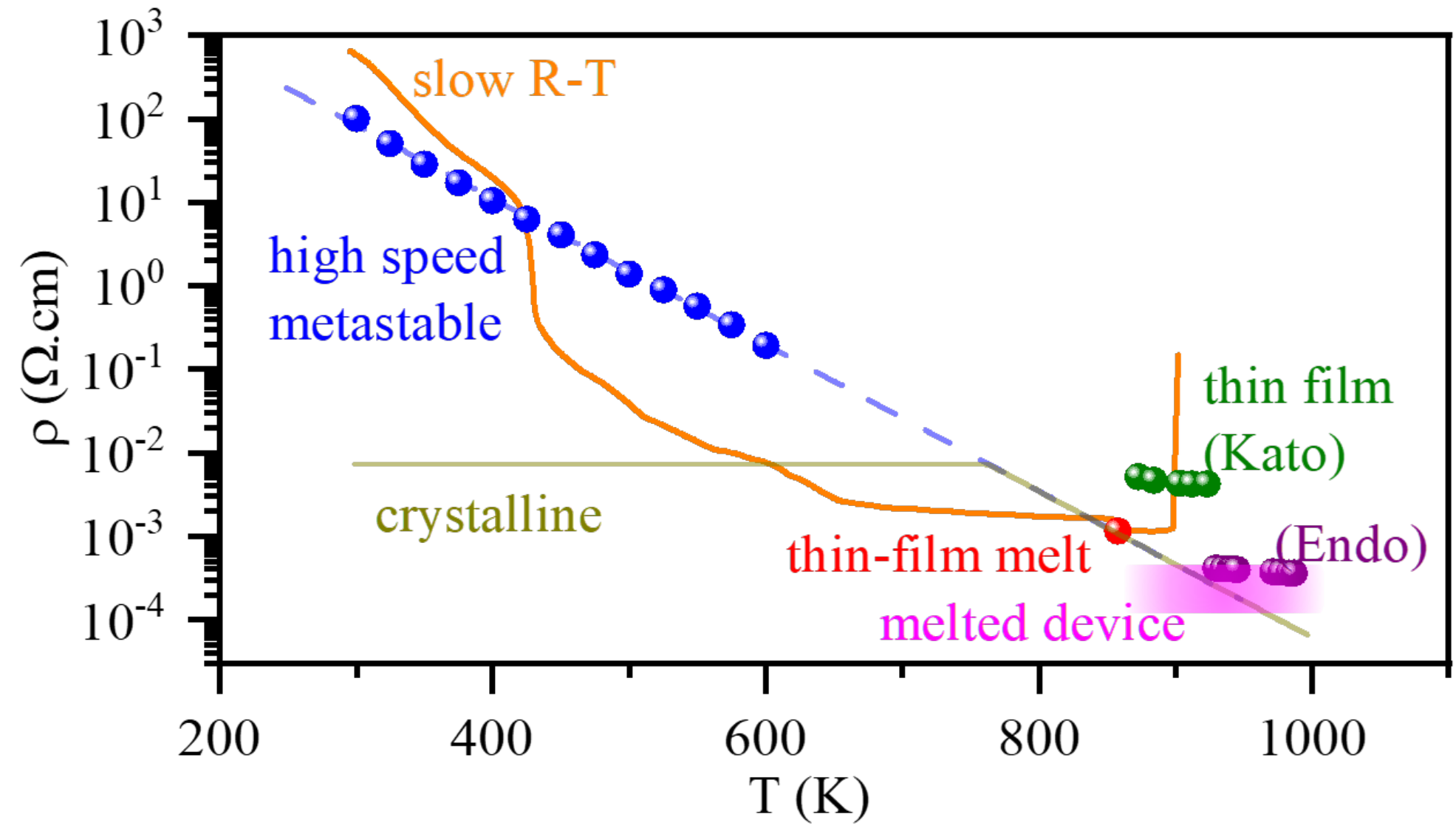}
\caption{Temperature dependent metastable resistivity $\rho$ (blue spheres) of amorphous GST measured at device level along with crystalline resistivity. Adapted from \cite{muneer2018activation}, licensed under CC BY 4.0.}
\label{fig:rho_vs_T.pdf}
\end{figure}

We use crystalline and metastable a-GST properties to study the cell behavior during high-speed cycling. Device-scale high-speed low-field resistivity versus temperature ($\rho$--$T$) measurements of metastable a-GST show a simple exponential response to temperature \cite{dirisaglik2015high} with a trend that matches the molten resistivity of GST at $T_{\mathrm{melt}}$ \cite{cil2013electrical} (Fig.~\ref{fig:rho_vs_T.pdf}). Assuming an Arrhenius behavior, we calculate the effective activation energy (\(E_A\)) as a function of temperature from the fit parameters $\rho_1 = 35137~\Omega\cdot\mathrm{cm}$ and $\alpha = 0.0202~\mathrm{K^{-1}}$, extracted from the experimental data \cite{hafiz2025modeling}, \cite{muneer2018activation} (Fig.~\ref{fig:rho_vs_T.pdf}):

\begin{equation}
\rho = \rho_1e^{- \alpha T} = \rho_0e^{\frac{E_A(T)}{k_BT}}
\end{equation}

Assuming $E_A(T_{\mathrm{melt}})=\frac{3}{2}k_B T_{\mathrm{melt}}$, we obtain:

\begin{equation}
\rho_0 = \rho_1e^{- \frac{3}{2} - \alpha T_{\mathrm{melt}}}
\end{equation}

From Eqs. (1) and (2), $E_A(T)$ can be written as:

\begin{equation}
\small
E_A(T) = k_BT\left\{ \frac{3}{2} + \alpha(T_{\mathrm{melt}} - T) \right\}
\end{equation}

$E_A(T)$ reaches zero at around $930~\mathrm{K}$ (Fig.~\ref{fig:rho_vs_T.pdf}), which we describe as the metal transition temperature ($T_{\mathrm{metal}}$) of GST.

\begin{figure}[!t]
\centering
\includegraphics[width=\columnwidth]{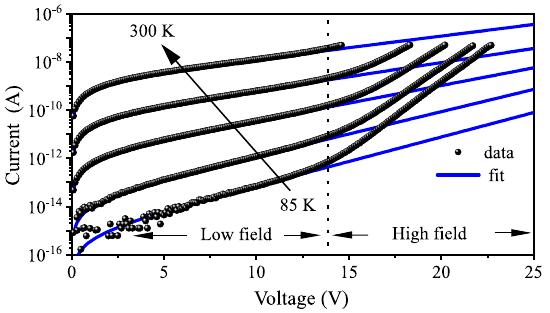}
\caption{I-V characteristics of stable amorphous GST line cells with $W\times L\times t\approx152~\mathrm{nm}\times710~\mathrm{nm}\times20~\mathrm{nm}$ measured at $85$, $125$, $175$, $225$, and $300~\mathrm{K}$. Adapted from \cite{khan2020accelerating}.}
\label{fig:IV_curves.pdf}
\end{figure}

\begin{figure}[!t]
\centering
\includegraphics[width=\columnwidth]{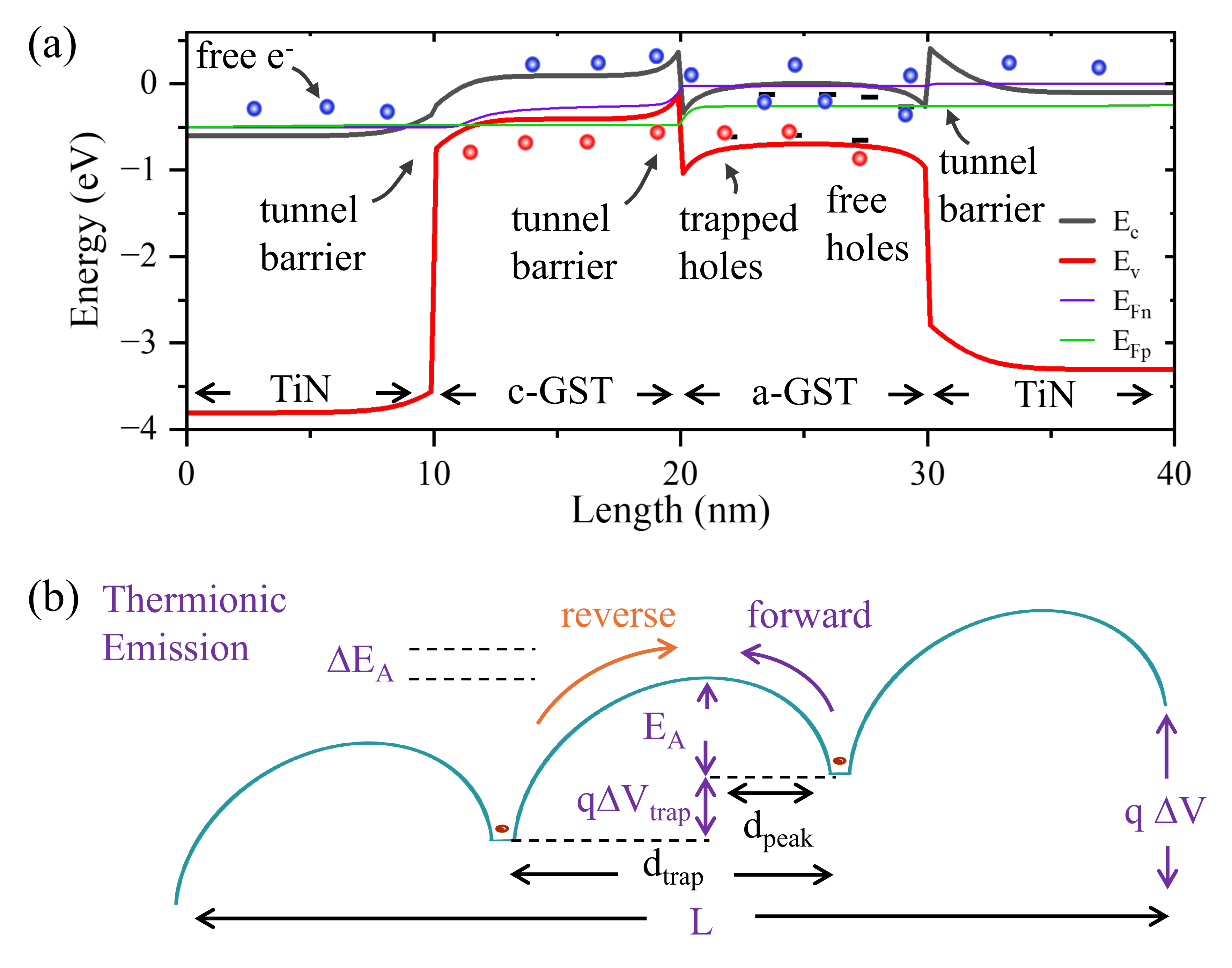}
\caption{(a) Room temperature band diagram of TiN/c-GST/a-GST /TiN structure under a mild bias of $0.5~\mathrm{V}$ for $20~\mathrm{nm}$ GST length with illustrative charge traps. The de-trapped holes cannot escape the potential well formed, and only electrons can contribute to transport. (b) Schematic conduction-band edge along transport direction. The trapped charges activation energy ($E_A$) is the peak energy barrier at zero applied bias minus the modulation by the applied bias $\Delta E_A(V)$ which depends on the ratio of the peak location ($d_{\mathrm{peak}}$) and trap separation ($d_{\mathrm{trap}}$).}
\label{fig:band_diagram.pdf}
\end{figure}

Temperature-dependent ($85$ to $300~\mathrm{K}$) high-field ($0$ to $\sim 50~\mathrm{MV/m}$) I-V sweeps performed on a-GST line-cells show distinct low-field ($<20~\mathrm{MV/m}$) and high-field ($>20~\mathrm{MV/m}$) responses \cite{khan2020accelerating}. High-field stress significantly accelerates resistance drift, and the devices stabilize within a few minutes. The subsequent I-V sweeps show stable device characteristics (Fig.~\ref{fig:IV_curves.pdf}) \cite{khan2020accelerating}. The low-field characteristics show hyperbolic sine behavior, which can be attributed to trap assisted transport of electrons (Fig.~\ref{fig:band_diagram.pdf}). The net current can be written as the difference between the forward and reverse currents that describe thermionic emission over an energy barrier modulated by the applied voltage \cite{tashfiq2022stopping}:
\begin{equation}
\begin{aligned}
I(V,T) &= I_{\mathrm{forward}} - I_{\mathrm{reverse}} \\
&= I_0 e^{-\frac{E_A - b\omega V k_B T}{k_B T}}
 - I_0 e^{-\frac{E_A - (b-1)\omega V k_B T}{k_B T}}
\end{aligned}
\label{eq:current_transport}
\end{equation}
where $I_0$ is a current scaling factor, $b=d_{\mathrm{peak}}/d_{\mathrm{trap}}$ and $\omega=qd_{\mathrm{trap}}/(k_BTL)$ are parameters extracted from fitting Eq. (4) to low-field experimental data in Fig.~\ref{fig:IV_curves.pdf}. $L$ is the length of the measured GST line cell, $d_{\mathrm{trap}}$ is trap-to-trap distance, and $d_{\mathrm{peak}}$ is the trap to barrier peak distance for forward transmission (Fig.~\ref{fig:band_diagram.pdf}). The activation energy ($E_A$) is defined as the peak barrier height when no bias is applied. We observe $b$ to decrease as a function of temperature and approach $\sim 0.5$ for $T>300~\mathrm{K}$ \cite{tashfiq2022stopping}, indicating that the barrier peak is approximately at the center position between the two traps. Expressed in terms of current density $J$, the model for $T>300~\mathrm{K}$ simplifies to:
\begin{equation}
\small
J(V,T)=
\frac{I(V,T)}{\mathrm{cross\mbox{-}section~area}}
=J_0 e^{-\frac{E_A}{k_B T}}
\left(e^{\frac{\omega V}{2}}-e^{-\frac{\omega V}{2}}\right).
\label{eq:J_transport}
\end{equation}

Taylor series expansion of Eq. 5 yields:

\begin{equation}
\small
\begin{aligned}
J(V,T) ={}& J_0 e^{-\frac{E_A}{k_B T}}
\left[
\left\{1+\frac{\omega V}{2}
+\frac{(\omega V/2)^2}{2}+\cdots\right\}\right.\\
&\left.
-\left\{1-\frac{\omega V}{2}
+\frac{(-\omega V/2)^2}{2}+\cdots\right\}
\right]
\end{aligned}
\label{eq:taylor_expansion}
\end{equation}
 
The low-voltage behavior can be modeled by neglecting the third and higher order terms and expressed in terms of electric field, $E=V/L$, as:
\begin{equation}
J(V,T)=J_0 e^{-\frac{E_A}{k_B T}}\omega V
=J_0 e^{-\frac{E_A}{k_B T}}\omega L E
=\frac{1}{\rho}E .
\label{eq:J_low_field}
\end{equation}
Hence, the electrical resistivity and \(J_0\) can be formulated as:

\begin{equation}
\rho = \rho_0e^{\frac{E_A}{k_BT}} = \frac{1}{\omega LJ_0}e^{\frac{E_A}{k_BT}}
\end{equation}

where $J_0$ is the current density for the no-barrier-for-forward-transmission case, $E_A/(k_BT)-\omega V/2=0$, which signifies the end of the transport limited by thermionic emission: using the $\rho(T)$ and $E_A(T)$ extracted from the metastable a-GST measurements (Fig.~\ref{fig:rho_vs_T.pdf}) and $\omega$ extracted from the stable measurements (Fig.~\ref{fig:IV_curves.pdf}), we calculate $J_0=188~\mathrm{MA/cm^2}$, which corresponds to a maximum velocity ($J_0/qn$) of $\sim 10^5~\mathrm{cm/s}$ for metastable a-GST, assuming carrier concentration $n=10^{22}~\mathrm{cm^{-3}}$ \cite{tashfiq2022stopping}. As a comparison, carrier saturation velocity in amorphous silicon is $1.7\times 10^4~\mathrm{cm/s}$ \cite{lin1995amorphous} and in crystalline silicon is $1\times 10^7~\mathrm{cm/s}$ \cite{becker2010measurements}.

The electronic transport in crystalline (fcc) GST is linear in nature ($J=E/\rho$) with $\rho(300~\mathrm{K})=6.9\times10^{-5}~\Omega\cdot\mathrm{m}$ \cite{adnane2016electrical} \cite{woods2017modeling2} (Fig.~\ref{fig:rho_vs_T.pdf}).

We model the thermal transport during device operation as:

\begin{equation}
\resizebox{\columnwidth}{!}{$\displaystyle
\underbrace{dC_p\frac{dT}{dt}}_{\substack{\text{Heat} \\ \text{absorbed}}} - \underbrace{\nabla \cdot (k\nabla T)}_{\substack{\text{Fourier} \\ \text{conduction}}} = - \underbrace{\nabla \cdot (J \cdot E)}_{\text{Joule heat}} - \underbrace{\nabla \cdot (JST)}_{\substack{\text{Thermoelectric} \\ \text{heat}}} + \underbrace{Q_H}_{\substack{\text{Latent heat of} \\ \text{phase change}}}
$}
\label{eq:heat transfer}
\end{equation}

where $d$ is the mass density, $C_p$ is the specific heat, $Q_H$ is the latent heat of phase change, $S$ is the Seebeck coefficient \cite{woods2017modeling2} and $k$ is the thermal conductivity, $0.27$ and $0.4~\mathrm{W\,m^{-1}\,K^{-1}}$ at $300~\mathrm{K}$ for amorphous and crystalline GST respectively \cite{lee2013phonon}. $S$ and $Q_H$ are temperature dependent. Modeling of $Q_H$ can be found in \cite{scoggin2018modeling}. The thermal transport model is solved along with electronic transport and phase change physics model \cite{scoggin2019modeling} in a fully coupled manner to capture the dynamics of device operation. Electric potential, current density, temperature, and nucleation-growth-amorphization of the GST crystal grains are calculated dynamically as the simulation progresses. A rate equation is used to track crystallinity, which can be generally described as:
\begin{equation}
\frac{\partial C_D}{\partial t} = \mathrm{Nucleation}+\mathrm{Growth}+\mathrm{Amorphization}
\end{equation}

where the nucleation term randomly generates nuclei depending on the local temperature. The growth and amorphization terms account for the increase and decrease in crystal grain sizes respectively. A single growth velocity versus temperature function, with a negative growth velocity for $T>T_{\mathrm{melt}}$, is used to model crystallization-amorphization velocity at the grain boundaries \cite{scoggin2019modeling}.

\begin{figure}[!t]
\centering
\includegraphics[width=\columnwidth]{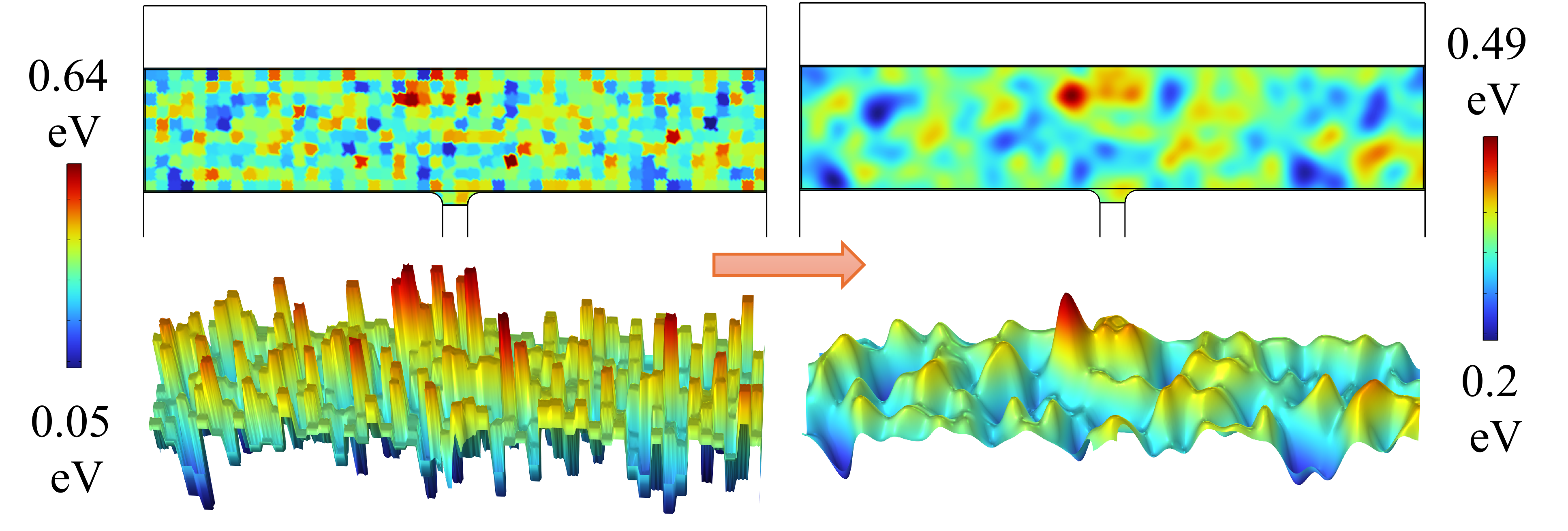}
\caption{Random $E_A$ map with $2~\mathrm{nm}\times2~\mathrm{nm}$ granularity in a $100~\mathrm{nm}\times20~\mathrm{nm}$ GST region (left) and the diffused $E_A$ used for simulations which reduces the $E_A$ gradients to enable convergence (right). The heater width is $4~\mathrm{nm}$.}
\label{fig:EA_map.pdf}
\end{figure}

We model variations in a-GST, which may be due to variations in composition, defects and trapped charges, by randomly varying the $E_A(T)$ extracted from the experimental studies using a random distribution function across the GST domain (Fig.~\ref{fig:EA_map.pdf}). In the studies we present here, we have used 30\% standard deviation in $E_A(T)$ in a 2D GST domain with $2~\mathrm{nm}\times2~\mathrm{nm}$ pixel size that is smoothed using a diffusion function to reduce the numerical complexity in the simulations.

\begin{equation}
\frac{\partial E_A}{\partial t} + \nabla\cdot \left( - c\nabla E_A \right) = 0
\end{equation}

\section{2D Modeling Results}
\begin{figure}[!t]
\centering
\includegraphics[width=\columnwidth]{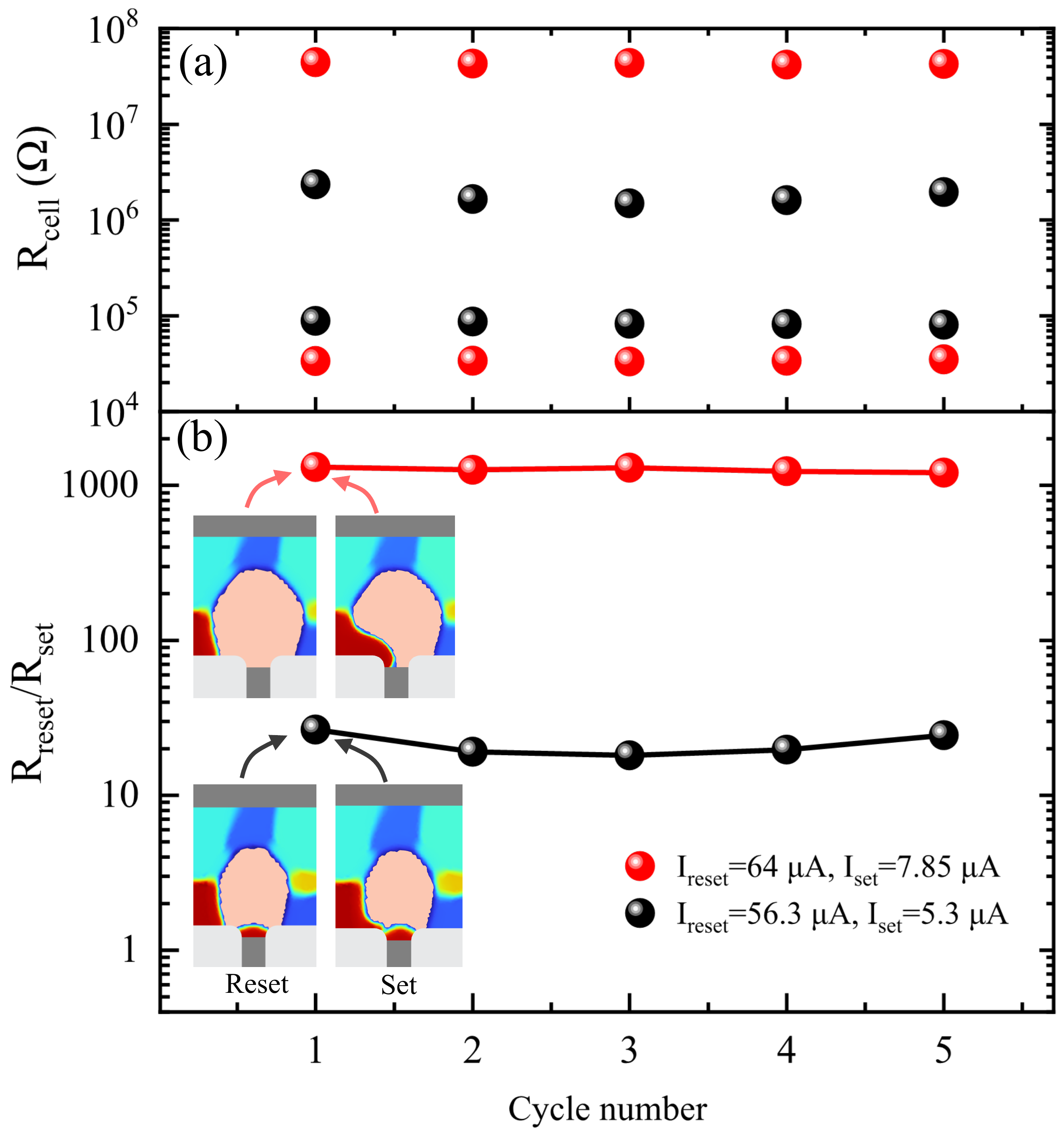}
\caption{(a) Cell resistances after Reset ($R_{\mathrm{reset}}$) and Set ($R_{\mathrm{set}}$) for five cycles. (b) A small increase in $I_{\mathrm{reset}}$ and $I_{\mathrm{Set}}$ results in a large increase in $R_{\mathrm{reset}}/R_{\mathrm{set}}$.}
\label{fig:memory_window_VTop.pdf}
\end{figure}

\normalcolor
Mushroom cells are typically fabricated as vertical cells with titanium nitride (TiN) contacts. Reset is performed with short and high amplitude pulses that melt GST in the programming region. Rapid termination of the pulse, quenching, results in amorphization. Set is typically performed using longer duration pulses, maintaining $T>T_{\mathrm{glass}}$ in the programming volume, to enable crystallization. Read is performed using small ($\sim 0.1~\mathrm{V}$) short duration pulses that do not disturb the state of the cell. The resistance contrast after Reset and after Set operations, $R_{\mathrm{reset}}/R_{\mathrm{set}}$, is referred to here as the memory window (Fig.~\ref{fig:memory_window_VTop.pdf}).

\begin{figure}[!tbp]
\centering
\includegraphics[width=\columnwidth]{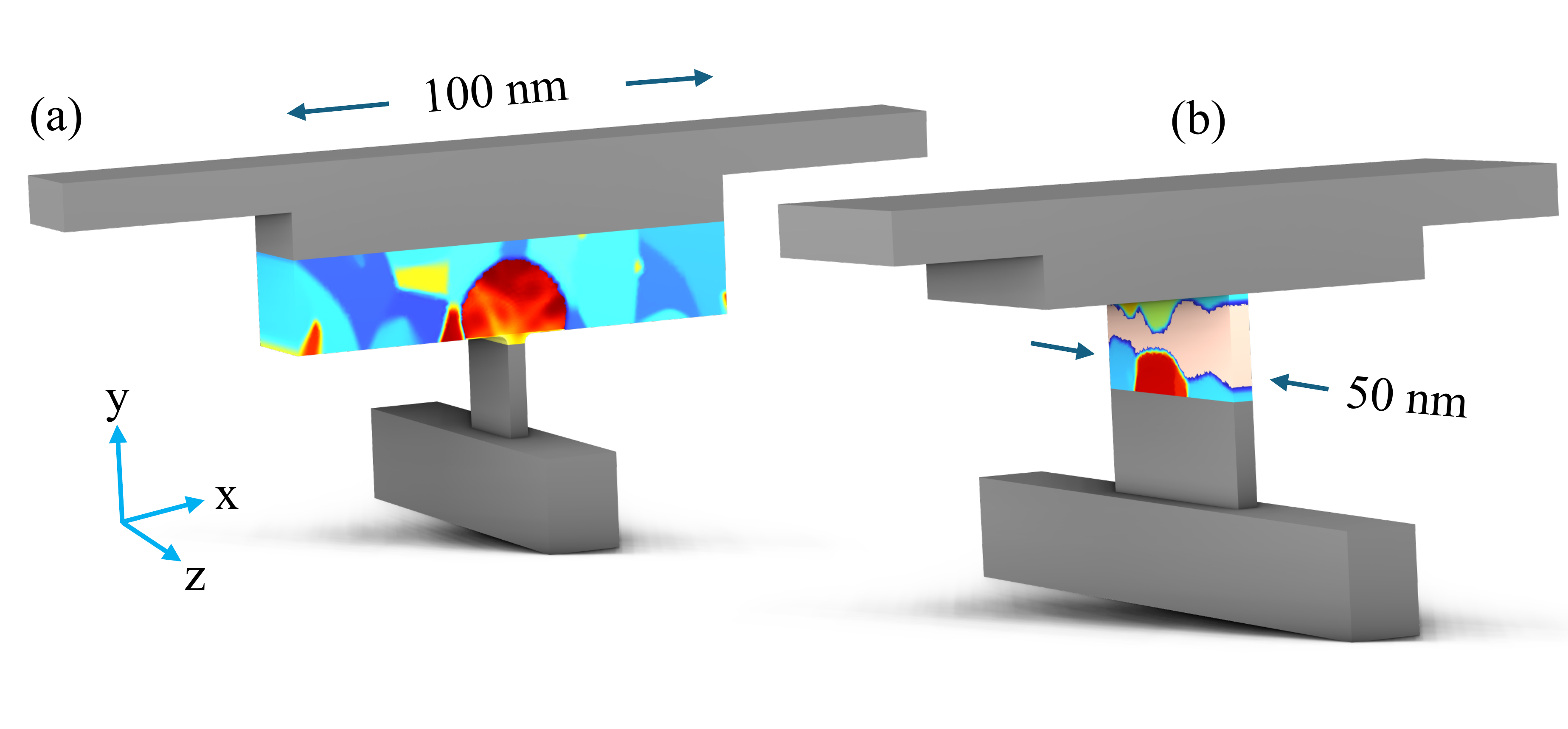}
\caption{3D extruded view of the mushroom cell for (a) x-y plane with $10~\mathrm{nm}$ out-of-plane depth ($z$ direction) and (b) out of mushroom-plane ($y$--$z$) with $4~\mathrm{nm}$ out-of-plane depth ($x$ direction) 2D simulations. (a) shows the current density inside the amorphous region.}
\label{fig:3D_structure.pdf}
\end{figure}

As 3D simulations are computationally expensive, we performed 2D geometry simulations for two cross-sections: typical 2D mushroom-cell simulation ($x$--$y$ plane) with $10~\mathrm{nm}$ out-of-plane depth, and out of mushroom-plane ($y$--$z$ plane) simulations with $4~\mathrm{nm}$ out-of-plane depth (Fig.~\ref{fig:3D_structure.pdf}). We choose the applied voltages to keep the access transistor in linear mode during Read and in saturation mode during programming (Set, Reset) operations (Fig.~\ref{fig:Reset_set_V,I,T_vs_t_VTop_on_VBot.pdf}).

\begin{figure}[!t]
\centering
\includegraphics[width=\columnwidth]{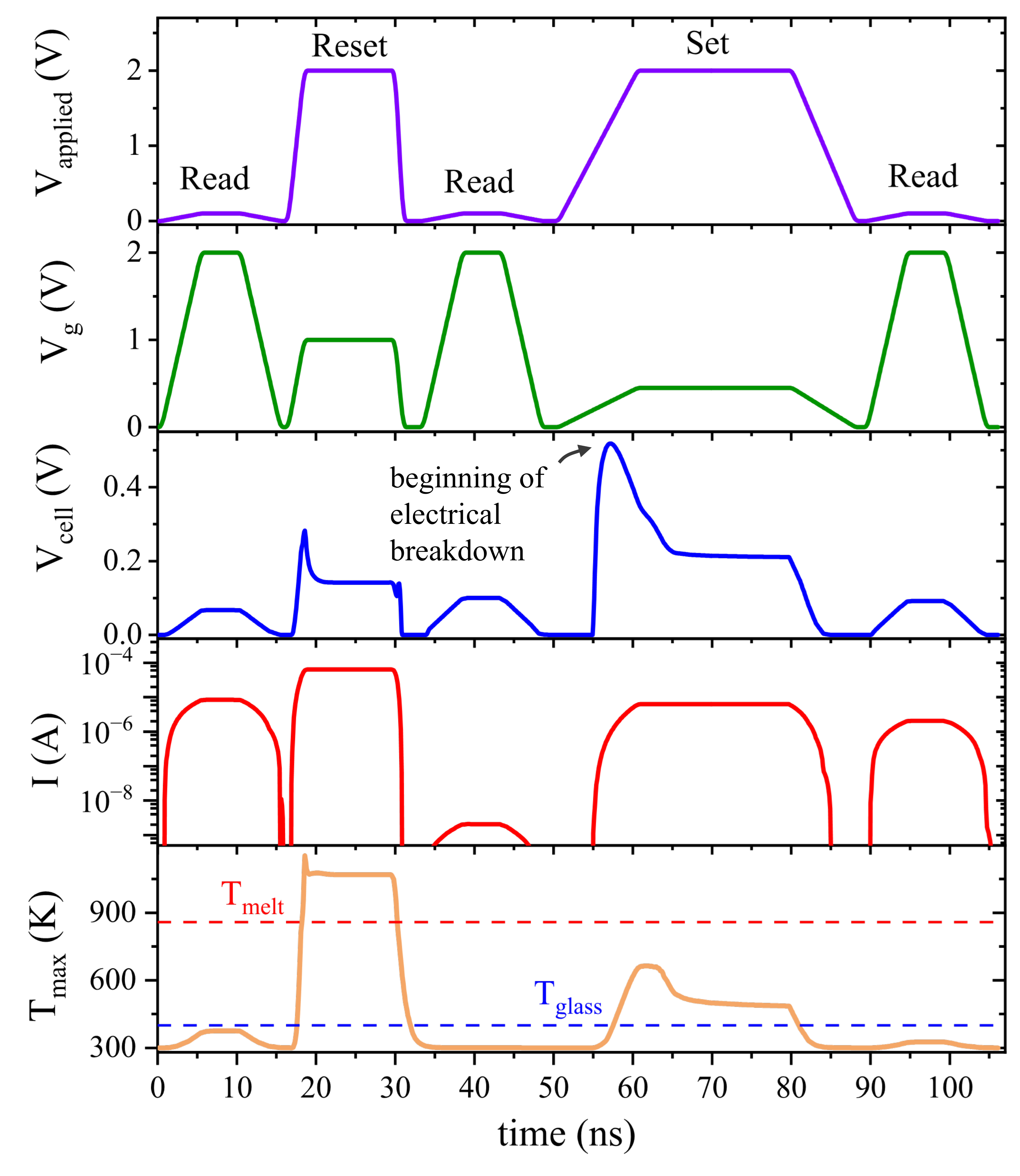}
\caption{Reset, Set and read pulses voltages ($V_{\mathrm{applied}}$, $V_g$ and $V_{\mathrm{cell}}$) and current. Set pulse has longer fall time compared to Reset pulse. PCM cells go through higher temperature in Reset than Set operation. Reading is performed with $V_g=2~\mathrm{V}$ to keep the n-MOSFET at low resistance, and $V_{\mathrm{applied}}$ =$0.1~\mathrm{V}$ not to disturb the state of the cell.}
\label{fig:Reset_set_V,I,T_vs_t_VTop_on_VBot.pdf}
\end{figure}

\begin{figure}[!t]
\centering
\includegraphics[width=\columnwidth]{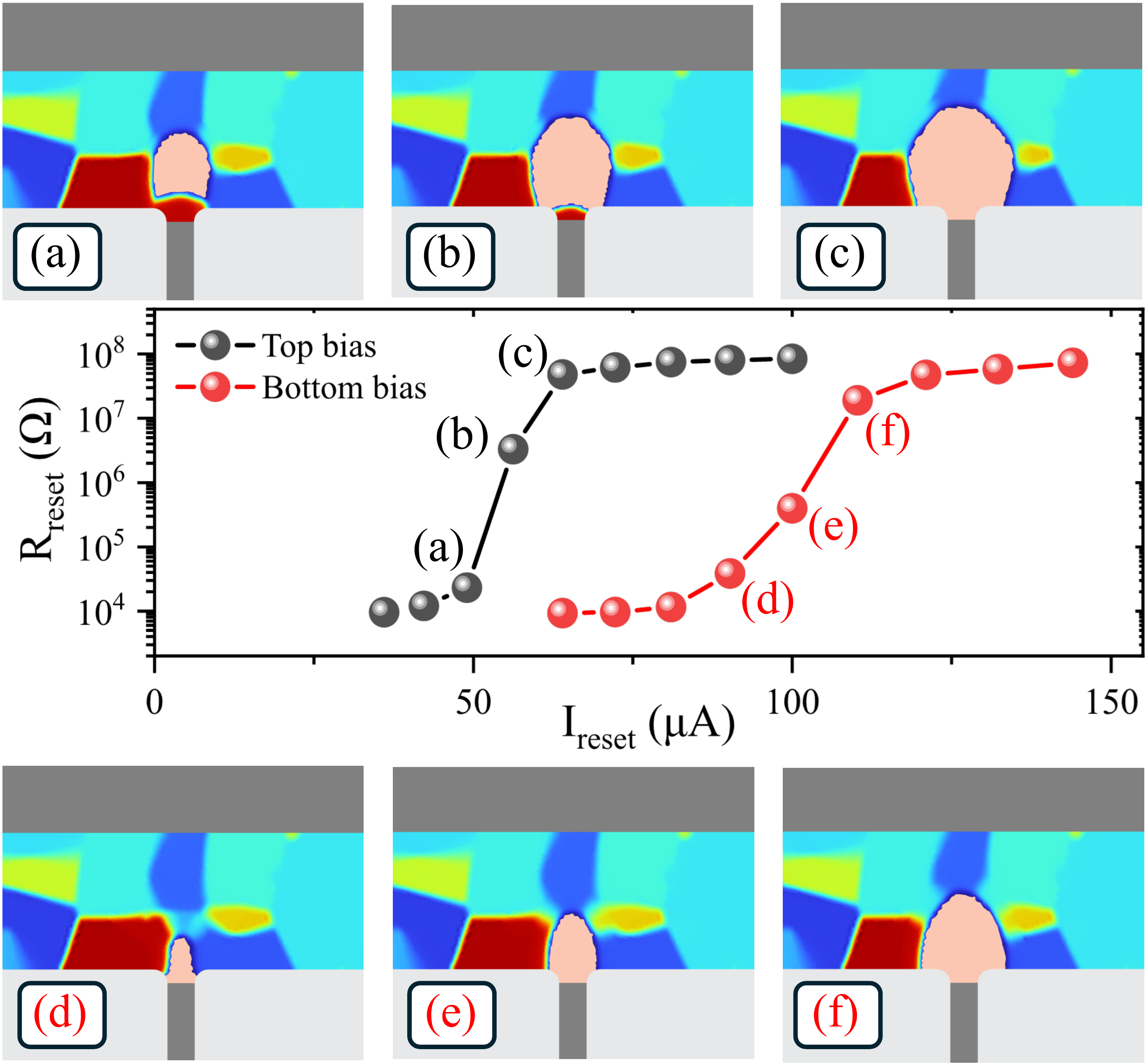}
\caption{Cell resistance after Reset operation versus Reset current for top and bottom bias polarities. (a-f) show Reset cell crystallinity profiles (cropped figures) for different Reset currents.}
\label{fig:Rreset_vs_Ireset.pdf}
\end{figure}

\subsection{\textcolor{subsectioncolor}{Reset Operation}}
\normalcolor

\begin{figure}[!tbp]
\centering
\includegraphics[width=\columnwidth]{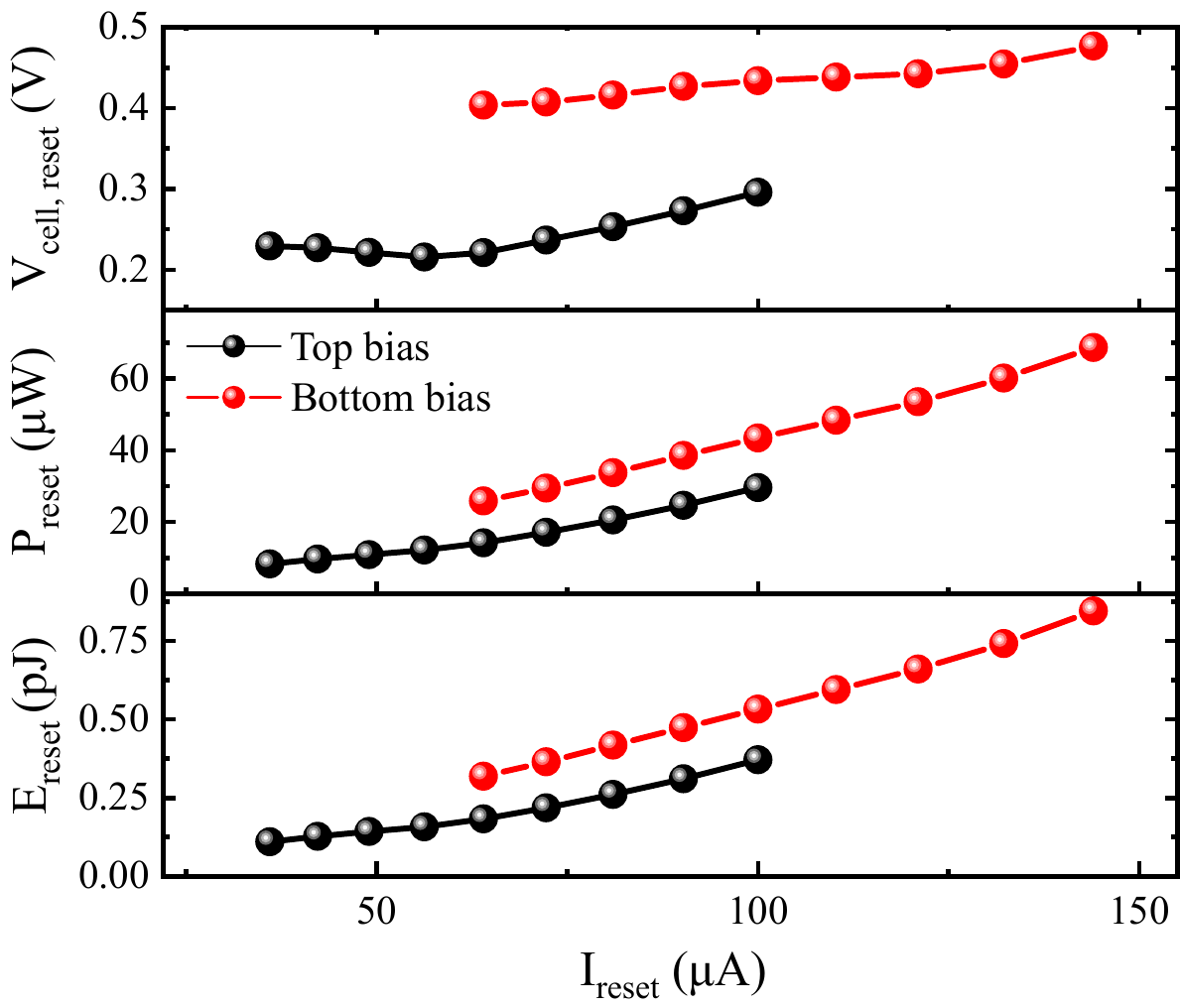}
\caption{PCM cell peak Reset voltage and power and Reset energy versus Reset current. It shows top bias polarity requires $\sim 2\times$ less voltage and current than the opposite polarity to achieve same $R_{\mathrm{reset}}$ ($\sim 48~\mathrm{M}\Omega$).}
\label{fig:Vcell_P_W_vs_Ireset.pdf}
\end{figure}

\begin{figure}[!tbp]
\centering
\includegraphics[width=\columnwidth]{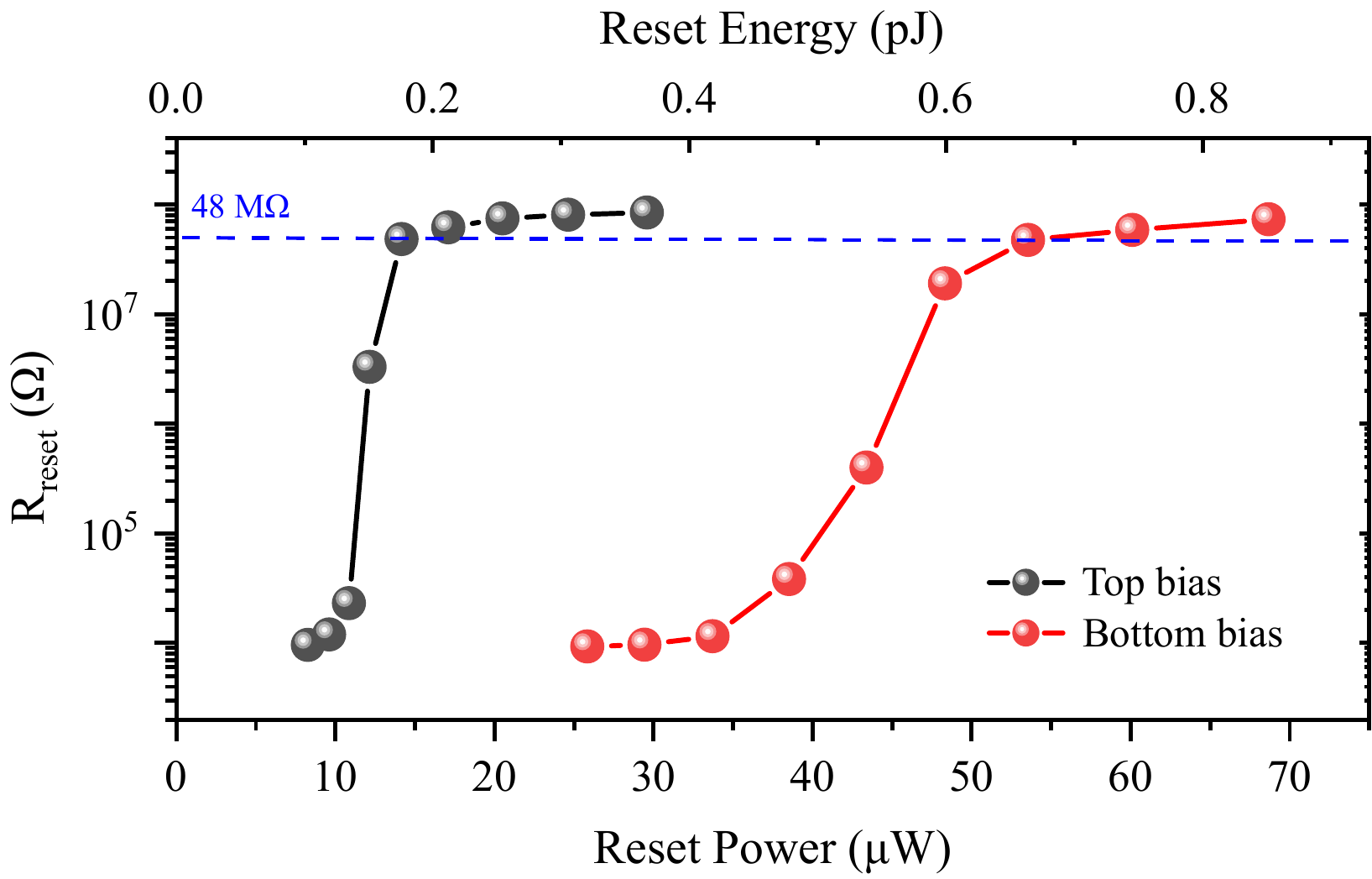}
\caption{$R_{\mathrm{reset}}$ versus peak Reset power and Reset energy for both polarities. Top bias Reset requires $\sim 3.5\times$ less power and energy than the bottom bias polarity to achieve same $R_{\mathrm{reset}}$ ($\sim 48~\mathrm{M}\Omega$).}
\label{fig:Rreset_vs_Power__Energy.pdf}
\end{figure}

Reset operation results in different amorphous volumes and resistances for different access transistor (nMOSFET) gate pulse amplitudes ($V_g$) at constant PCM electrode bias ($V_{\mathrm{applied}}$) (Fig.~\ref{fig:circuit.pdf}). Increasing $V_g$ of the nMOSFET leads to increased Reset current, $I_{\mathrm{reset}}$, which increases the size of the amorphous volume and memory window (Fig.~\ref{fig:memory_window_VTop.pdf}, Fig.~\ref{fig:Rreset_vs_Ireset.pdf}). $R_{\mathrm{reset}}$ significantly increases if an amorphous region completely covers the bottom contact. The increase in $R_{\mathrm{reset}}$ is relatively mild for higher $I_{\mathrm{reset}}$ beyond this level (Fig.~\ref{fig:Rreset_vs_Ireset.pdf}). Similar saturating behavior is observed for both top and bottom bias circuits with a significant difference in the required $I_{\mathrm{reset}}$, and Reset power and energy, $P_{\mathrm{reset}}$ and $E_{\mathrm{reset}}$, due to thermoelectric effects (Fig.~\ref{fig:Vcell_P_W_vs_Ireset.pdf}, Fig.~\ref{fig:Rreset_vs_Power__Energy.pdf}). The top-bias condition successfully resets at lower currents \cite{ciocchini2015impact,faraclas2014modeling,kashem2022digital} (Fig.~\ref{fig:Vcell_P_W_vs_Ireset.pdf}). Higher power and energy can be delivered to the cell by increasing $I_{\mathrm{reset}}$ (Fig.~\ref{fig:Vcell_P_W_vs_Ireset.pdf}). However, over-reset is not desirable since power consumption and required access device size (limiting maximum $I_{\mathrm{reset}}$) are major concerns for PCM design.

The thermoelectric term in Eq. \ref{eq:heat transfer}, \(\nabla\cdot (JST)\), which can be written in terms of its electron and hole contributions as \(\nabla\cdot (JST) = \ \nabla\cdot \left( J_{n}S_{n}T \right) + \nabla\cdot \left( J_{p}S_{p}T \right)\), is very significant for Reset and Set operations due to (i) asymmetric cell structure, (ii) extreme current density and thermal gradients, and (iii) substantial generation and annihilation rate of electrons and holes at (or close-proximity to) the material and solid-liquid interfaces \cite{bakan2013high}. Fcc-GST and TiN are degenerate p-type and n-type semiconductors. The interface of these two materials is expected to experience substantial Peltier heating or cooling due to carrier recombination or generation, depending on the current polarity. Conduction in a-GST is expected to be due to free electrons due to band alignment and the inability to inject holes from the TiN or fcc-GST into the valance band of a-GST. Therefore, the TiN/a-GST interface is expected to have very small Peltier heating or cooling and the a-GST/fcc-GST interface is expected to have large Peltier heating or cooling. Molten GST is a very narrow band-gap semiconductor with very large electron and hole concentrations (n, p). We estimate $n_{\mathrm{liquid}}=p_{\mathrm{liquid}}\sim 1\times10^{22}~\mathrm{cm^{-3}}$ from the latent heat of fusion: latent heat of fusion / pair = kinetic energy of the electron + kinetic energy of the hole + band gap at $T_{\mathrm{melt}}$ = $3kT/2+3kT/2+3kT/2$.

\begin{figure}[!tbp]
\centering
\includegraphics[width=\columnwidth]{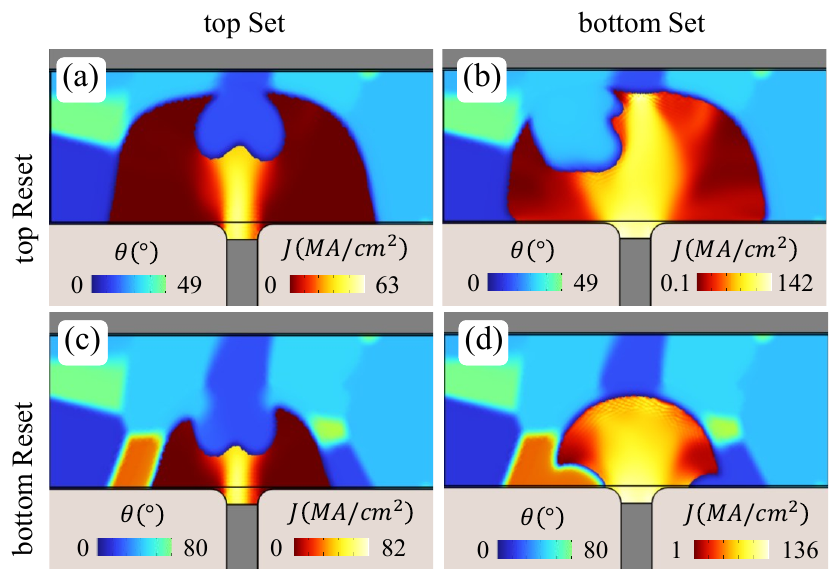}
\caption{Crystal orientation angles for fcc-GST and current density profiles for a-GST during top and bottom polarity Set of cells Reset with top and bottom polarity, without accounting for thermal boundary resistances.}
\label{fig:Without_TBR.pdf}
\end{figure}

\begin{figure}[!tbp]
\centering
\includegraphics[width=\columnwidth]{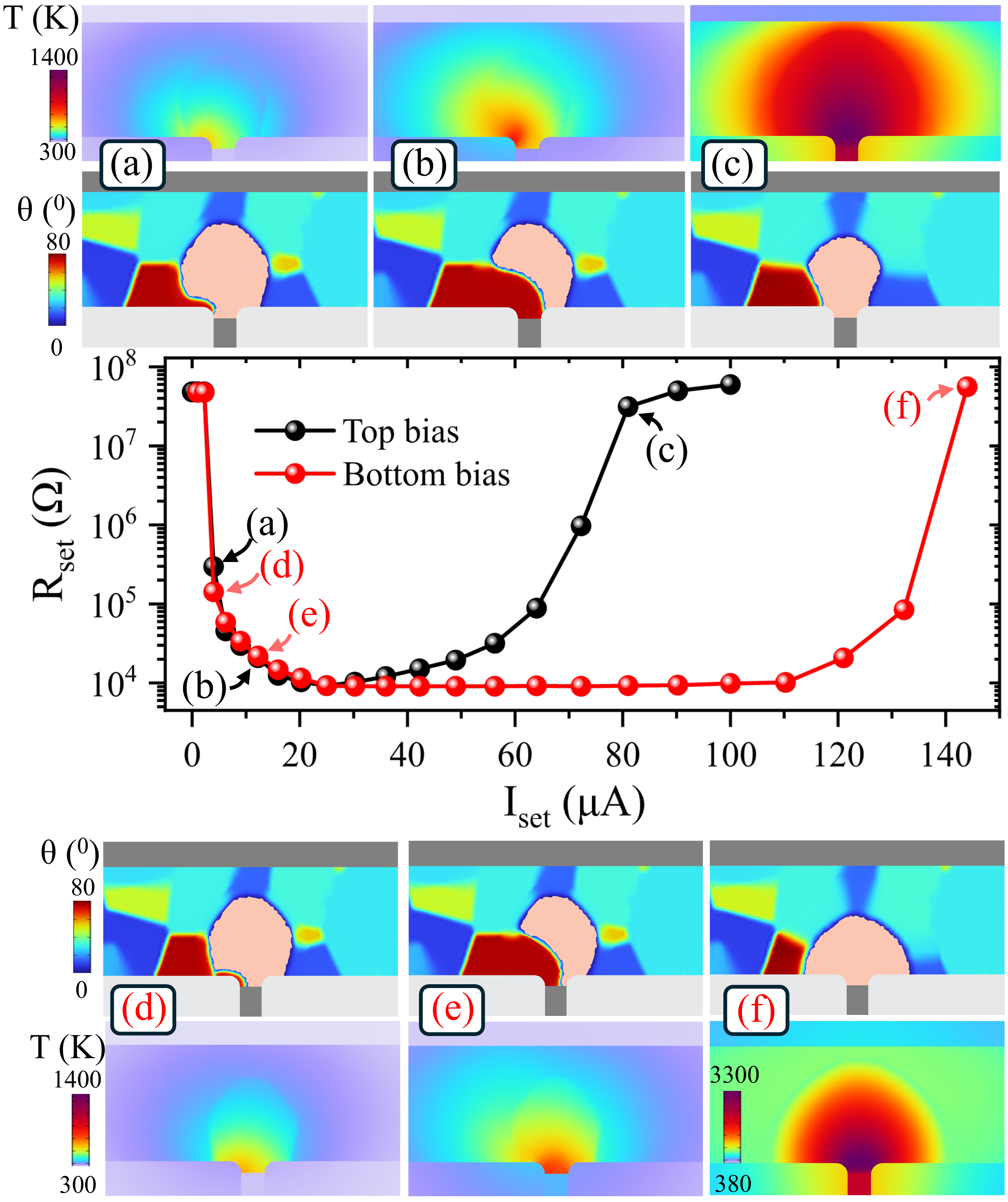}
\caption{Cell Resistance after Set operation versus Set current for top and bottom bias polarities. (a-f) Set cell crystalline orientation and peak temperature profiles for different Set currents. The cell was previously Reset with top bias polarity with $64~\mu\mathrm{A}$. $W_{\mathrm{heater}}=4~\mathrm{nm}$.}
\label{fig:Rset_vs_Iset_small_mushroom_VTop_reset}
\end{figure}

\subsection{\textcolor{subsectioncolor}{Set Operation}}
\normalcolor

Once the PCM cell is Reset, a subsequent Set pulse is applied to crystallize the GST (Fig.~\ref{fig:Reset_set_V,I,T_vs_t_VTop_on_VBot.pdf}). $I_{\mathrm{set}}$ is controlled by $V_g$. The simulated cells, which start with identical reset condition (top-bias), can be set with $5~\mu\mathrm{A}<I_{\mathrm{set}}<64~\mu\mathrm{A}$ for top-bias and $5~\mu\mathrm{A}<I_{\mathrm{set}}<132~\mu\mathrm{A}$ for bottom-bias conditions (Fig.~\ref{fig:Rset_vs_Iset_small_mushroom_VTop_reset}). Excessive $I_{\mathrm{set}}$ results in melting and re-amorphization of the programming region. The bottom bias polarity has larger range for Set operation due to thermoelectric cooling at the interface between fcc-GST and the bottom TiN contact. The location of the dominant filament is a significant function of the thermal boundary conditions and the thermal boundary resistance (TBR) \cite{lee2013phonon2,lee2010thermal,reifenberg2010thermal} between TiN and GST. Crystallization starts from one side of the mushroom, close to the bottom GST/SiO$_2$ interface. If the simulations are conducted without TBR at the material interfaces, wider mushrooms are formed and the crystallization path depends on both the polarity and the shape of the mushroom (Fig.~\ref{fig:Without_TBR.pdf}).

\begin{figure}[!h]
\centering
\includegraphics[width=\columnwidth]{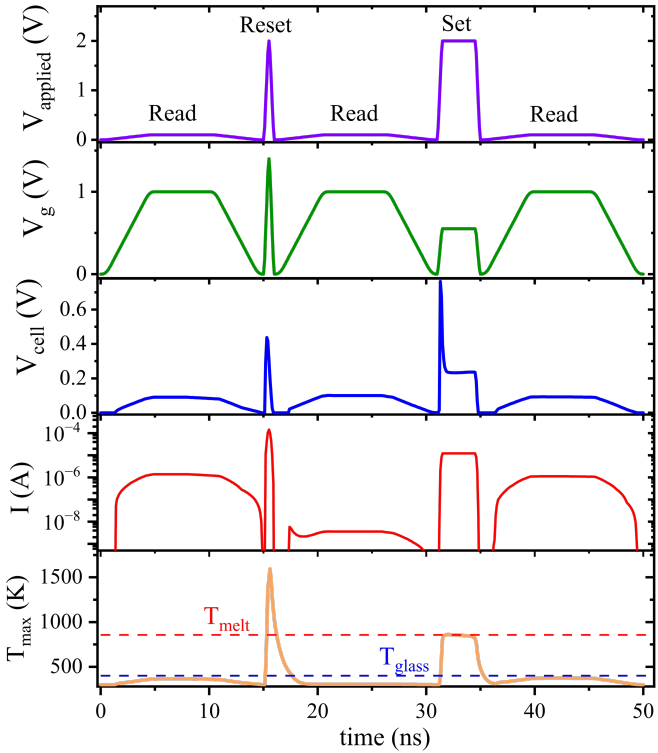}
\caption{Shorter Reset ($\sim 1~\mathrm{ns}$) and Set pulses ($\sim 4~\mathrm{ns}$) voltage and current waveforms used for the out-of-plane simulations. $V_{\mathrm{applied}}$ is kept constant, and only $V_g$ is varied for Set/Reset operations.}
\label{fig:Reset_set_V,I,T_vs_t.pdf}
\end{figure}

\begin{figure}[!tbp]
\centering
\includegraphics[width=\columnwidth]{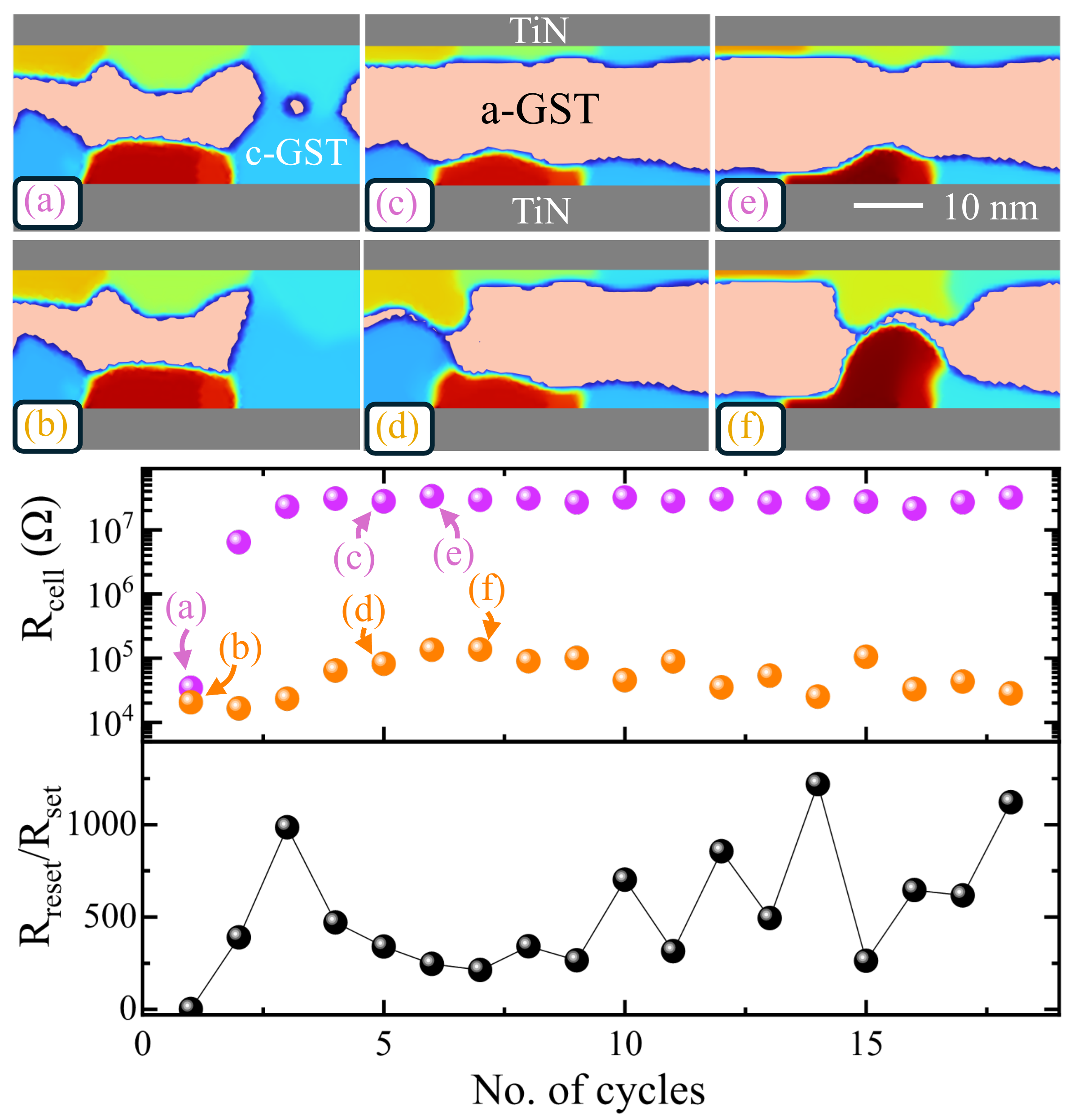}
\caption{Cell resistances after Reset ($R_{\mathrm{reset}}$) and Set ($R_{\mathrm{set}}$) operation for 18 cycles (shown in Fig.~\ref{fig:Reset_set_V,I,T_vs_t.pdf}). The first Reset/Set cycle does not switch the cell but leads to successful switching starting from the next cycle. $R_{\mathrm{reset}}/R_{\mathrm{set}}$ cycle-to-cycle variability is due to randomness in phase change dynamics and filamentary transport in a-GST.}
\label{fig:Rcell__Rreset_Rset_vs_cycles.pdf}
\end{figure}

\subsection{\textcolor{subsectioncolor}{Orthogonal-Plane ($y$--$z$) Simulations}} 
\normalcolor
We performed 2D simulations on the ($y$--$z$) plane in Fig.~\ref{fig:3D_structure.pdf}(b), which is orthogonal to the mushroom ($x$--$y$) plane, and reduced the Reset ($\sim 1~\mathrm{ns}$) and Set ($\sim 4~\mathrm{ns}$) pulse durations to observe the role of filamentary conduction and variability along the depth of the cell (Fig.~\ref{fig:Rcell__Rreset_Rset_vs_cycles.pdf}). The initial Reset pulse resulted in amorphization of a $\sim 35~\mathrm{nm}$ section of the $50~\mathrm{nm}$ depth of the cell, which is not a successful Reset. The second Reset pulse resulted in amorphization of the crystalline region that still bridged the two contacts, successfully resetting the cell. The subsequent Set and Reset pulses, completed in $5~\mathrm{ns}$ and $4~\mathrm{ns}$, respectively, including the time to cool down to room temperature, form and break a bridge that is $\sim 10~\mathrm{nm}$ in depth, leading to a clear ($>100\times$) memory window (Fig.~\ref{fig:Rcell__Rreset_Rset_vs_cycles.pdf}). This small region, which is approximately $4~\mathrm{nm}\times10~\mathrm{nm}\times10~\mathrm{nm}$ ($x$--$y$--$z$), functions as the programming volume rather than the full semi-cylindrical amorphized mushroom region ($\sim 10~\mathrm{nm}\times10~\mathrm{nm}\times50~\mathrm{nm}$). Hence, 2D mushroom simulations with out-of-plane depth of $10~\mathrm{nm}$ (Fig.~\ref{fig:3D_structure.pdf}(a)) are expected to capture the operation dynamics of cells with larger out-of-plane depth.

Scaling of the out-of-plane depth is not expected to reduce the required $I_{\mathrm{reset}}$ and $I_{\mathrm{set}}$, unless scaled below $10~\mathrm{nm}$, but can improve cell-to-cell and cycle-to-cycle variability. Larger out-of-plane depths can provide increased reliability as the cells would be more immune to over-reset: if the programming region is over-reset, current can still be flown at a different location to continue cycling of the cell (Fig.~\ref{fig:Rcell__Rreset_Rset_vs_cycles.pdf}). While this increases variability, it will also increase the lifetime and cycling endurance of the cells.

\section{Conclusions}
\normalcolor
We investigated thermoelectric effects and filamentary transport in mushroom phase change memory cells during Reset and Set operations using an electrothermal finite element framework that accounts for temperature-dependent transport parameters, phase change dynamics, and stochastic activation energy distribution in the amorphous phase. We observe a $2\times$ reduction in reset current and $3.5\times$ reduction in energy consumption and peak Reset power with top-polarity bias condition due to thermoelectric effects. Our model predicts a significant change in required Set current as a function of mushroom shape and size, but no clear advantage for either bias polarity. Thermal boundary conditions and especially thermal boundary resistances at the GST/TiN interface at the bottom contact play a significant role in Reset and Set dynamics, mushroom shape, current and peak power requirements. Filamentary conduction in a-GST during Set operation initiates local heating and phase change and determines the location, shape and volume of the programming region, which may be significantly smaller than the semi-cylindrical mushroom region. In the case of relatively wide amorphized mushrooms, we observe crystallization starting from the top of the mushroom for the top-bias Set and from the side of the mushroom for bottom-bias Set. Out-of-plane simulations capture cycle-to-cycle variability and show that cells with larger out-of-plane depth are expected to show increased variability but also increased lifetime and cycling endurance. These findings highlight the central role of filamentary transport and thermoelectric effects in PCM switching dynamics and energy consumption and contribute to the broader effort of advancing PCM toward scalable, low-power, and reliable memory technologies for future computing applications.

\section*{Acknowledgment}
The authors thank F. Dirisaglik for resistivity measurements, S. Muneer for activation energy extraction, and Z. Woods for constructing the phase change physics model. 


\end{document}